\documentstyle[11pt,epsfig]{article}

\def\bea{\begin{eqnarray}}
\def\eea{\end{eqnarray}}
\def\be{\begin{equation}}
\def\ee{\end{equation}}

\begin{document}

\title{\bf QCD}

\author{{\sc D. Espriu}\thanks{E-mail: espriu@ecm.ub.es}\\
Departament d'Estructura i Constituents de la Mat\`eria and IFAE,\\
Universitat de Barcelona,\\
Diagonal 647,
E-08028 Barcelona.}

\date{}

\maketitle

\begin{abstract}This is the written version of the lecture on
deep inelastic scattering and related topics
in  QCD,
delivered
in the course of the XXVI International
Meeting on Fundamental Physics to an audience of young
experimentalists.
The aim is fundamentally pedagogical. I review the
theoretical setting of the Altarelli-Parisi equations, discuss
recent determinations of $\alpha_s$ from
deep-inelastic scattering and then move to the
kinematical region explored by HERA. In the way I mention some
unsolved theoretical problems. I discuss low-$x$
physics and to what extent $\log\frac{1}{x}$ resummations are called for.
\end{abstract}

\vfill
\vbox{UB-ECM-PF-99/03\null\par
January 1999}

\clearpage

\section{Introduction}
These notes are not an
introduction to
Quantum Chromodynamics (QCD), the theory
of strong interactions. Many excellent textbooks exist
where the interested reader can find clear expositions
of the subject\cite{qcd}. In fact we shall assume here that
the reader is familiar with the basic technical tenets
of perturbative QCD, such as Feynman diagrams, dimensional
regularization, renormalization, beta function and so on.
We pretend rather to give physical insight
of the reasons behind a rather remarkable theoretical
development which is nearly as old as QCD itself, namely
deep inelastic scattering. Why is it so remarkable?

QCD is, in a way, a rather simple theory (specially when
compared to the intricacies of the electroweak part of the
Standard Model). It is just a simple extension of good old
Quantum Electrodynamics. Instead of matter fields carrying
electrical charge  +1 (and  anticharge -1), if we are talking
about electrons, the matter fields of QCD (the quarks)
carry a new quantum
number: color. Color can take three different values (and their
corresponding anti-values). Furthermore the intermediate bosons
(the gluons), unlike photons which cannot change
the charge of a particle, change the color of a quark. They may for
instance turn a red quark into a blue quark. This simple fact
implies that the gauge group of QCD is much larger that the
$U(1)$ of QED. Since every quark comes in three copies, whose labels
get exchanged, there is a $SU(3)$ invariance\footnote{The reason why
the symmetry group is $SU(3)$ and not $U(3)$ ---which would have
nine gluons--- has to do with the mathematical fact that
$\epsilon_{\alpha\beta\gamma}$ is not an invariant tensor
of $U(3)$. This tensor is necessary to build combinations
of three quarks which are antisymmetric, as required by
Fermi statistics. For instance, $\Delta^{++}
= \vert u^\uparrow u^\uparrow u^\uparrow\rangle $. Being
the lightest hadron with this quark contents we expect to have
the three quarks in the ground state, hence in a symmetric wave function.
This is in contradiction with Fermi statistics. The contradiction can be
solved if we admit the existence of a new quantum number $\alpha$ and
$\vert u^\uparrow u^\uparrow u^\uparrow\rangle
= {1\over \sqrt{6}}\epsilon^{\alpha\beta\gamma}
\vert u_\alpha^\uparrow u_\beta^\uparrow u_\gamma^\uparrow\rangle$.}.

Simple as this theory may seem, it is not an easy
matter in QCD to relate theory and experiment.
It is well known that the
fields and particles  we know how to compute with
(with the simplest tool at our disposal, perturbation
theory) are not those that are observed by experimentalists
in their detectors due to the phenomenon of confinement.
Quarks and gluons
are real, but they cannot be detected as free particles
as they are known
to be confined inside hadrons. In view of this is quite remarkable
that there are  theoretical techniques enabling us to put the
theory
to very stringent tests.

The phenomenon of confinement comes about because the coupling
constant of QCD (which is relatively small at large values
of the momentum transfer, the  value quoted by the Particle Data
Group\cite{pdg} is
$\alpha_s(M_Z)=0.119\pm 0.002$; notice the truly amazing precision,
which may soon be reduced to a mere 1\%) becomes strong as the
energy decreases. The behaviour predicted by the
renormalization-group is, at one loop,
\be
\alpha_s(Q)={-{\pi}\over{{\beta_1\over_2}
  \log(Q^2/\Lambda_{QCD}^2)}}.\label{alfa}
\ee
where $\Lambda_{QCD}$ is a renormalization-group invariant,
but scheme dependent, quantity and $\beta_1= -11/2 +N_f/3$. The preferred
value
is $\Lambda_{QCD}=219^{+25}_{-23}$ MeV (5 flavours, $\overline{MS}$-scheme).
The meaning of the scale-dependent, renormalized
coupling constant is roughly the following: it is the ``effective"
coupling, relevant at the scale $Q$, namely, the one that (within the
choosen scheme) minimizes further quantum corrections, in particular
resumming all large logs.
From  (\ref{alfa}) we see that precisely at the
scale $Q^2=\Lambda_{QCD}^2$, the effective coupling has a pole. Of course
well before that scale is reached perturbation theory becomes
completely unreliable, and the $1/r$ potential of the
perturbative interaction change to a stronger behaviour, possibly
to a linear $\sim r $ behaviour.
When computing the production of any physical
hadron (typically of  mass $\sim$ few $\Lambda_{QCD}$), perturbation
theory is completely useless.

One instance where
perturbative QCD can be applied is to inclusive processes, provided
that the characteristic momentum transfer is large enough.
These will not be discussed in this lecture. The interested reader
can look at the
determination of $\alpha_s$
through $R_{had}$
$R_\tau$, for instance in.\cite{R}

A clear application of perturbative
QCD is  provided by deep inelastic scattering. The subject is
now over twenty years old and, by now, perturbative QCD has
been tested to a high degree. Furthermore, the commissioning
of HERA has opened a new kinematical region where it may be
possible to study the onset of non-perturbative effects in a
controlled fashion.
The exploration of this region is a fascinating subject
interesting on its own right.

Due to the lack of time and space we have not included two sections that, in
our view, should be in any general review of perturbative QCD and deep
inelastic scattering. The first one
concerns the
so-called ``spin of the proton" problem\cite{proton}.
Another topic that is not
covered at all is the study of the photon structure
functions.\cite{foton}
The list of references is very incomplete and those provided  merely reflect
personal tastes.

\section{Logs in QCD}
Beyond tree level most Feynman diagrams
are ultraviolet divergent. Take for instance the
one diagram contributing at one loop
to the gluon propagator.
Neglecting external momenta, the integral over the
momenta of the internal particles is of the form
\be
\int {{d^4k}\over{(2\pi)^4}}{{k^\alpha k^\beta}
\over k^4}=\infty . \label{integral}
\ee
To make sense of the theory  and get a finite result we must introduce
a cut-off $\Lambda$ and counterterms.
A possible method is to perform a subtraction at some
$q^2=-\mu^2$. For instance, for the self-energy
of the gluon propagator
\be
\Pi(q^2)-\Pi(-\mu^2)\equiv\Pi_R(q^2)={\rm finite}.\label{Pi}
\ee

Alternatively we can make sense of the integrals
using dimensional regularization by continuing
the dimensionality from 4 to $n=4+2\epsilon,$
\be
\int {{d^4k}\over{(2\pi)^4}}\to\int{{d^nk}\over{(2\pi)^n}},\label{reg}
\ee
and  subtract just the poles in $1/\epsilon$
(minimal subtraction, $MS$) or also the $\gamma_E-\log 4\pi$
that always accompanies the singularity in
$1/\epsilon$ (improved minimal subtraction, $\overline{MS}$).
For instance, for the
quark contribution
to the gluon self-energy, one has the following
result after computing the integral in $n=4+2\epsilon$
dimensions
\be
\Pi(q^2)=-{\alpha_s\over {6\pi}}\delta_{ab}(
{1\over\epsilon}+\gamma_E+\log{{m^2}\over{ 4\pi\mu^2}}+\dots).
 \label{renorm}
\ee
Using the $MS$ and $\overline{MS}$ schemes one gets
\be
\Pi_{MS}(q^2)=-{\alpha_s\over {6\pi}}\delta_{ab}(
\gamma_E+\log{{m^2}\over{ 4\pi\mu^2}}+\dots)
\qquad
\Pi_{\overline{MS}} (q^2)=-{\alpha_s\over {6\pi}}\delta_{ab}(
\log{{m^2}\over{\mu^2}}+\dots)\label{renorm1}
\ee

The above expressions illustrate the appeareance of ultraviolet
logs $\log{q^2/\mu^2}$ through the renormalization procedure. There are,
however, other source of logs in QCD. They are of the
form
$\log{q^2/ \lambda^2}$
where $\lambda^2$ can is some external momentum squared or a small
(mass)$^2$ that we have given by hand to the (massless) gluon. While the
former are
associated to ultraviolet
divergent integrals (integrals with a bad behaviour
when the internal momentum is large), the latter
are infrared logs
 and are related to Feynman diagrams
with a bad behaviour when one or more external
momenta vanish.
Ultraviolet logs
appear in any renormalizable field theory after
renormalization. On the contrary, infrared logs
appear whenever a theory has massless particles in
the spectrum (such as photons or gluons).
A given Feyman diagram can give rise to both
type of singularities at the same time.

\begin{figure}[h]
\epsfysize=2cm
\epsfbox{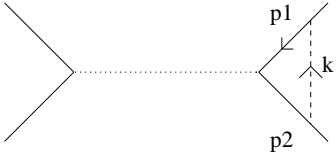}
\caption{ Example of diagram leading to an infrared
divergence. \label{fig1}}
\end{figure}

There actually two classes of infrared logs caused
by massless particles. The so-called
infrared divergences arise from the
presence of a {\it soft massless} particle
($k^\mu\to 0$). For instance in the
process $e^+e^-\to \mu^+\mu^-$ at the
one loop level
we have to compute the integral  (figure \ref{fig1})

\begin{figure}[h]
\epsfysize=1.3cm
\epsfbox{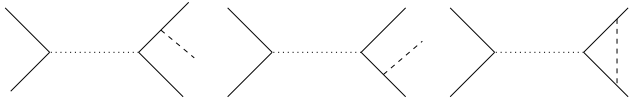}
\caption{ Real and virtual photons have to be included for IR
safe results. \label{fig2}}
\end{figure}

\be
\int {{d^4k}\over {(2\pi)^4}}
{1\over { k^2 [(p_1+k)^2-m^2][(p_2+k)^2-m^2]}}.
\ee
When $p_1^2=p_2^2=m^2$ the integral
behaves for $k^\mu\to 0$ as
\be
\int{{d^4k}\over {(2\pi)^4}}{1\over k^4}.\label{sing}
\ee
and diverges. This divergence is
unphysical so it must be cancelled by something
else. The Bloch-Nordsieck theorem\cite{bloch} states that
in inclusive enough cross-sections the
infrared logs cancel. What do we mean by `inclusive
enough'?  A detector will not be able to discern
a `true' muon from a muon accompanied by a soft
enough photon (with $\vec{k}\to 0$). Therefore,
we have
to consider diagrams where a soft photon is
radiated by the muon, square the modulus of the
amplitude and integrate over the available phase
space (which actually depends on the experimental cut). When
this is done the result is infrared finite. The
relevant diagrams are depicted in figure \ref{fig2}.

The other type of infrared logs are called mass singularities.
They occur in theories with massless particles
because two {\it parallel massless} particles have an
invariant mass equal to zero
\be
k^2=(k_1+k_2)^2=\Vert (\omega_1+\omega_2,0,0,
\omega_1+\omega_2)\Vert=0.\label{invar}
\ee
The appeareance of such a mass singularity is
illustrated in figure 3

\begin{figure}[h]
\epsfysize=2.5cm
\epsfbox{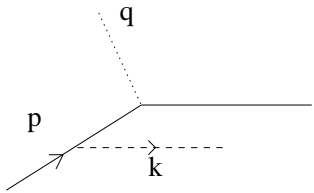}
\caption{Diagram with a mass singularity.
\label{fig3}}
\end{figure}

\be
{1\over {(p-k)^2}}={1\over {p^2+k^2-2k^0p^0+
2k^0p^0\cos \theta}},\label{propa}
\ee
the denominator vanishes when we set all particles
on shell ($p^2=k^2=0$) and $\theta\to 0$
(i.e. $\vec{k}$ is parallel to $\vec{p}$). Even if one of the two particles
is massive there is a singularity, provided the
3-momenta are parallel.

The Kinoshita-Lee-Nauenberg theorem\cite{kln} ensures
that for inclusive enough cross section the mass
singularities also cancel. Both for mass singularities
and for infrared divergences there is a trade-off
between $\lambda^2$, the infrared regulator of a massless particle,
and the energy and angle resolution of the
inclusive cross section $\Delta E$, $\Delta \theta$.

In practice, it is better to regulate the infrared logs using
dimensional regularization (introducing $\lambda^2$ leads to
difficulties with gauge invariance). Real gluon emission
diagrams are regulated by performing the phase space integration
in $n$ dimensions.

There is in fact a lot of physical insight
hidden in the infrared logs.
Physical arguments
tell us that the probability of finding a `bare' isolated
muon should be {\it zero}. We know this because
detectors are unable to tell apart a muon from
a muon plus one soft photon or indeed from a
muon plus any number of soft photons.
Infrared divergences in QED can be summed up and
then one sees that the probability of finding
an isolated muon is indeed zero and not infinite as
the one loop diagram led us to believe. Whenever
a Feynman diagram is infrared divergent it means
that we have forgotten something relevant.

Let us consider in QED the interaction of a charged
fermion with an external source  and let us expand
in the number of {\it virtual} photons $n$
(i.e. in the number of loops). The total
amplitude will be expressed
as
\be
M(p,p^\prime)=\sum_{n=0}^\infty M_n(p,p^\prime)
\label{expon}
\ee
then a calculation shows that
\bea
	   M_0 &=&m_0, \nonumber \\
	   M_1 &=&m_0\alpha B+m_1,\nonumber \\
	   M_2 &=&m_0{{(\alpha B)^2}\over 2}+m_1\alpha B + m_2,\nonumber \\
	       &&\dots \label{exxx}
\eea
The quantities $m_n$ are IR-finite, while $B$ is IR-divergent.
The series in (\ref{expon}) can be summed up
\be
M=\exp(\alpha B)\sum_{n=0}^\infty m_n,\qquad m_n\sim\alpha^n,
\label{expon1}
\ee
and
$B$ can be obtained just from the lowest order diagram. Introducing
an IR cut-off $\lambda$,
$B\sim -\log {m^2/ \lambda^2}$,
which indeed shows that when we remove the cut-off the probability
of finding an isolated charged fermion is zero
in QED. The addition
of soft photons changes that result, multiplying the
total amplitude by a factor $\sim (\Delta E/\lambda)^2$.
There is a trade between the infrared regulator
and $\Delta E, \Delta \theta$. The latter are, of course,
detector-dependent quantities.

Although only partial results exist\cite{expinqcd}, it is believed that
a similar exponentiation takes place in QCD. Due to the
confinement subtleties it is unclear whether the
suppression factor is compensated by radiation
of soft gluons. Even if this compensation does actually take
place that would not disprove confinement, only that
confinement would have nothing to do with the
structure of infrared singularities of the theory.

The previous discussion can be summarized in the
following way: due to IR
singularities one is forced to consider
cross sections not of individual
particles in the final state, but rather of
bunches of particles, each `hard' quark
and gluon surrounded by a `soft' cloud of
gluons and, perhaps, quarks.
These bunches are called `jets'.

The Bloch-Nordsieck and Kinoshita-Lee-Nauenberg
theorems guarantee the finiteness of the
cross-sections. We have to define
an energy and angle resolution. For
instance, if $p$ is the momentum of a
primary quark
we can impose that the energy of each
soft particle in its jet satisfies
$k^0_i<\epsilon p_0$ and also that
$\arg (\vec{p},\vec{k}_i)<\delta$.
We will get singularities
of the form $\alpha_s\log\epsilon\log\delta$
when $\epsilon,\delta\to 0$. The specific details depend
on the precise definition of the jet.

\section{Free Parton Model}
The counterpart of having an effective coupling constant which grows
at low energies is that the theory becomes simple at high
energies, making perturbative calculations
possible (at least {\it some} of them). Indeed a brilliant confirmation
of the existence of nearly free constituents
inside the nucleon was provided more than
twenty years ago by a series of experiments
carried out at SLAC\cite{slac}. Then it became possible
to scatter electrons off nucleons in fixed
target experiments with a typical momentum transfer
$\sim 1 - 10$ (GeV)$^2$, a kinematical range unexplored until that time.
The
kinematics of Deep Inelastic Scattering (DIS) processes
is shown in fig. 4.

\begin{figure}[h]
\epsfysize=2.5cm
\epsfbox{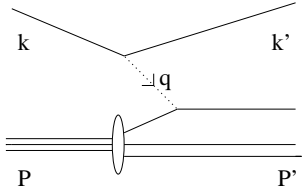}
\caption{The kinematics of deep inelastic scattering.
\label{fig4}}
\end{figure}

The virtual intermediate boson
is far off its mass-shell and
scatters off a quark or gluon in a time
of ${\cal O}({1/{\sqrt{-q^2}}})$. Typically
quarks and gluons are themselves off-shell by an
amount of ${\cal O}(\Lambda_{QCD})$. After
the scattering the outgoing particles recombine
into hadrons in a time of ${\cal O}(1/\Lambda_{QCD})$.
Thus Deep Inelastic is a two-step process

\begin{itemize}
\item  Short distance scattering occurs with a large
momentum transfer. Well described by perturbation theory.

\item  Outgoing particles recombine. Not calculable
in perturbation theory.

\end{itemize}

However, the second part can be side-stepped all together
for fully inclusive rates. Then perturbation theory
is adecuate to describe many features of DIS.

If we place ourselves in the center of mass of the
hadron and virtual intermediate boson both particles
move very fast towards each other. Whatever components
the hadron contains they will all have
moments parallel to $P^\mu$, up to transversal
motion of ${\cal O}(\Lambda_{QCD})$. Let us write
\be
p^\mu=x P^\mu.\label{pmu}
\ee
The squared CM energy of the lepton and proton constituent
will be
\be
\hat{s}=(xP+k)^2\simeq 2xPk\simeq xs.\label{hat}
\ee
We
neglect masses (as well as the fact that constituents are
off-shell
by ${\cal O}(\Lambda_{QCD})$)
The final momentum of the constituent is $xP+q$.
Therefore
\be
0\simeq (xP+q)^2\simeq 2x Pq +q^2,\label{zero}
\ee
so $x=-q^2/2Pq$. If $\nu$ is the energy
transfer in the LAB system, we can also
write
\be
x={{-q^2}\over{2\nu m_N}}\label{defx}
\ee
$m_N$ being the nucleon mass. It is convenient to
introduce
\be
y={{Pq}\over {Pk}}=1-{{Pk^\prime}\over {P k}}\label{defy}
\ee
In the lab frame $y=\nu/E$ and $0\le y\le 1$. $y$ is thus
the relative energy loss of the colliding lepton.

Let us for the time being ignore altogether QCD
interactions and let us assume that constituents
of the nucleons (which we will call partons) are
free. DIS will then be described
by an incoherent sum over elementary processes. The partonic
differential cross sections
in the LAB frame will be

\medskip
\noindent
$\bullet$ $\nu q$, $\overline{\nu} q$-scattering

\be
{{d\hat\sigma_\nu}\over {dy}}
=({{g^2}\over {4\pi}})^2{{\pi m E}\over {(q^2-M_W^2)^2}}
[g_L^2+g_R^2(1-y)^2], \label{neut}
\ee
\be
{{d\hat\sigma_{\overline{\nu}}}\over {dy}}
=({{g^2}\over {4\pi}})^2{{\pi m E}\over {(q^2-M_W^2)^2}}
[g_R^2+g_L^2(1-y)^2]. \label{neut1}
\ee

\noindent
$\bullet$ $eq$-scattering

\be
{{d\hat\sigma_e}\over {dy}}
=Q^2 {{4\pi\alpha^2m E}\over {q^4}}[1+(1-y)^2].\label{elect}
\ee
The neutral current sector is dominated by $\gamma$ exchange
below $q^2=M_Z^2$, so we have not bothered to include $Z$
exchange. In (\ref{elect}) $Q$ is the quark electric charge (in units
of $e$) and $m$ is the
target mass. Since $p^\mu=x P^\mu$, we just
take $m=x m_N$. Then, for instance,
\be
{{d^2\hat\sigma_e}\over {dx dy}}
=Q^2 {{4\pi\alpha^2 xm_N E}\over {q^4}}[1+(1-y)^2].\label{elect1}
\ee

Let $u(x)dx, d(x)dx,...$ be the number of $u,d,...$
quarks with momentum fraction between $x$ and $x+dx$
in a nucleon. Then $xu(x), xd(x),...$ will be
the fraction of the nucleon momentum carried by
$u,d,...$ quarks. We, of course, identify  quarks
with partons and, since we assume that they are free, proceed to
sum incoherently over the different scattering
possibilities. For instance in
$ep\to eX$
\be
{{d^2\sigma}\over {dxdy}}= {{2\pi\alpha^2}\over s}
{{1+(1-y)^2}\over {xy^2}}
[{4\over 9}(u(x)+\overline{u}(x))+
{1\over 9}(d(x)+\overline{d}(x))+
{1\over 9}(s(x)+\overline{s}(x))].\label{elect2}
\ee
(We neglect here the possible contribution from the
sea of heavy quarks in the nucleon.) Other
DIS processes weigh differently quarks
and antiquarks. For instance, in $\nu p\to \mu X$
if $-q^2\ll M_W^2$ we have
\be
{{d^2\sigma}\over {dxdy}}= x {{G_F^2 s}\over
\pi}[ c_c^2 d(x)+s_c^2 s(x)+\overline{u}(x)(1-y)^2],
\label{nup}
\ee
with $c_c=\cos \theta_c$, $s_c=\sin \theta_c$, the
cosinus and sinus of the Cabibbo angle, respectively.

 The parton distribution functions (PDF) $q(x)$ are
quantities which are not
calculable within perturbative
QCD, as we will see.

Probably the first thing
that one learns is that gluons are very important.
From the SLAC-MIT data\cite{slac}
\be
Q=U+D+S=\int_0^1 dx x (u(x)+d(x)+s(x))\simeq 0.44,\label{sea}
\ee
\be
\bar{Q}=\bar{U}+\bar{D}+\bar{S}=
\int_0^1 dx x (\bar{u}(x)+\bar{d}(x)+\bar{s}(x))\simeq 0.07.
\label{sea1}
\ee
The total fraction of momentum carried by quarks (and
antiquarks) is only about 50\% The rest is carried by gluons (parametrized
by a PDF $g(x)$), showing that although the naive quark model works very
well is just a gross simplification as a model of hadrons, at least at
large $-q^2$. In fact we know the asymptotic values
of (\ref{sea})  and (\ref{sea1}) based in the equipartition of energy in
a free theory (since, asymptotically, QCD is free)
\be
\int_0^1 dx x q(x)\to \frac{3N_f}{16+3N_f}\qquad
\int_0^1 dx x g(x) \to \frac{16}{16+3N_f}. \label{limit}
\ee
From the above
limiting values we see
 and at higher energies the total
momentum carried by {\it constituent} or {\it valence} quarks
diminishes and that an equally important role is played
by particles from the Dirac sea of the nucleon.

Another example where the quark model fails to describe some
basic features of hadrons is provided by the `spin of the proton'
problem\cite{proton}. $\mu$-scattering on polarized targets shows that
the fraction of the total spin of the proton that can naively be
associated to constituent quarks is surprisingly small. We shall
not dwelve on this matter further here.

Nevertheless,
there are some obvious sum rules for the parton
distribution functions which can  ultimately be explained in terms
of the quark model. For the proton
\be
\int_0^1 dx (u(x)-\bar{u}(x))= 2,\label{sumrule1}
\ee
\be
\int_0^1 dx (d(x)-\bar{d}(x))= 1,\label{sumrule2}
\ee
\be
\int_0^1 dx (s(x)-\bar{s}(x))= 0.\label{sumrule3}
\ee
On QCD grounds we expect that this free parton model
description of the hadrons becomes more and more accurate
when $-q^2\to \infty$, $\nu\to\infty$, while
keeping $x$ fixed. This limit is known as
Bjorken scaling and in the strict $-q^2=\infty$ limit everything depends
just on $x$.

Let us now try to rederive the previous results in a
more theoretical setting. Let us consider
for instance $\nu p$ scattering. Then
\be
{{d^2\sigma}\over {d(-q^2) d\nu}}=
{{G_F^2 m_N}\over {\pi s^2}} L^{\mu\nu} H_{\mu\nu},
\label{ope1}
\ee
where
\be
L^{\mu\nu}={1\over 8}{\rm Tr}
[\gamma^\mu(1-\gamma_5)\gamma^\alpha\gamma^\nu
(1-\gamma_5)\gamma^\beta] k_\alpha k_\beta \label{ope2}
\ee
is the trace over the leptonic external lines, and $H_{\mu\nu}$
is given by
\be
\sum_{X}\langle P\vert J_\mu(0)\vert X(P^\prime)\rangle
\langle X(P^\prime)\vert J_\nu(0)\vert P\rangle
=\int d^4z e^{iqz}\langle P\vert J_\mu(z)J_\nu(0)\vert
P\rangle ,\label{gkw}
\ee
which is just
${\rm Im}\Pi_{\mu\nu}(q)$,
with
\be
\Pi_{\mu\nu}(q)=\int d^4z e^{iqz} \langle P\vert T J_\mu(z)J_\nu(0)\vert
P\rangle.\label{ope4}
\ee
We decompose $H_{\mu\nu}$ as
\be
H_{\mu\nu}=-g_{\mu\nu}F_1+{{P_\mu P_\nu}\over {\nu m_N}}F_2 +{i\over {2\nu
m_N}}\epsilon_{\mu\nu\rho\sigma}P^\rho q^\sigma F_3 \label{ope5}
\ee
(If we
assume that we are working with non-polarized targets $P$ and $q$ are the
only vectors at our disposal.) $F_1$, $F_2$ and $F_3$ are called the nucleon
structure functions. Using the kinematical relations $x=-q^2/2\nu m_N$ and
$y=2m_N\nu/ s$ we get
\be
d(-q^2) d\nu= \nu s dx dy,\label{ope6}
\ee
\be
{{d^2\sigma}\over {dx dy}}={{G_F^2 s}\over {2\pi}} [F_1 xy^2+
F_2(1-y)-F_3xy(1-{y\over 2})].\label{ope7}
\ee
Let us now compare with the free
parton model. We see that (restoring the $\nu p$ index, to make apparent that
the structure functions are process dependent)
\be
F_1^{\nu
p}(x)=c_c^2(\bar{u}(x)+d(x))+s^2_c(s(x)+\bar{u}(x)),\label{ident1}
\ee
\be
F_2^{\nu p}(x)=
2xc_c^2(\bar{u}(x)+d(x))+2xs^2_c(s(x)+\bar{u}(x)),\label{ident2}
\ee
\be
F_3^{\nu
p}(x)= 2c^2_c(\bar{u}(x)-d(x)) +2 s^2_c(-s(x)+\bar{u}(x)).\label{iden3}
\ee
For other
processes the actual expressions may vary but the structure functions are
always linear combinations of the parton distribution functions, i.e.
$F_2(x)=x\sum_f \delta_f q_f(x)$, etc.

Note that in the free parton model
\be
F_L(x)=F_2(x)-{{F_1(x)}\over {2x}}=0.\label{longi}
\ee
This is the
Callan-Gross relation, which actually is not an  exact one;
it gets modified when the $q^2$ dependence is
included, i.e. we depart from the strict $-q^2=\infty$ limit.

An exact sum rule, which is easily expressed in terms of the structure
function $F_2(x)$ was given by Adler
\be
\int_0^1 \frac{dx}{x}(F_2^{\bar{\nu} I} - F_2^{\nu I})=
4\langle I_3\rangle \label{adler}
\ee
where $I_3$ is the third component of the target of isospin $I$.
Other sum rules are not exact, except in the strict free parton model, but
their violations are computable within perturbative QCD. Two instances
are the Gross-Llewellyn-Smith sum rule
\be
\frac{1}{2}\int_0^1 dx (F_3^{\nu p}+F_3^{\bar{\nu} p}) =
\int_0^1 dx (u(x)-\bar{u}(x)+d(x)-\bar{d}(x))+{\cal O}(\alpha_s)=
3 + {\cal O}(\alpha_s),\label{GLS}
\ee
and the Gottfried sum rule
\be
\int_0^1 \frac{dx}{x}(F_2^{\mu p}-F_2^{\mu n})= \frac{1}{3}
+{\cal O}(\alpha_s) +\int_0^1 dx (\bar{u}(x)-\bar{d}(x)).\label{gott}
\ee
Violations to the different sum rules are a theoretically clean way
to extract $\alpha_s$. Of course in practice things are
difficult because the sum rules involve an integral over
all values of $x$, which is always poorly known
in some range of the integrand and extrapolations are needed.

If we assume that $\alpha_s$ is extracted from some other
source, the Gottfried sum rule provides some interesting
information on the sea contents of light antiquarks
in the nucleon. The collaboration NMC\cite{NMC} has determined that
for the proton
\be
\bar{U}-\bar{D}\equiv\int_0^1 dx (\bar{u}(x)-\bar{d}(x))= -0.15\pm 0.036.
\label{isospin}
\ee
In the proton quark sea there are many more $d$-type antiquarks
than $u$-type. This is in fact a large isospin violation, much
larger than expected to the mass difference of the quarks
and, in fact, goes, at least naively, in the opposite
direction.

The isospin violation is larger at low values of $x$ ($x\le 0.2$)
and also at low values of $Q^2$, which hints that long-distance physics
(quite remote from perturbative QCD) is called for. The enhancement
of $d$-type antiquarks is confirmed by other
experiments. For instance Na51\cite{Na51}  finds at $x=0.18$
$\bar{u}/\bar{d}=0.51\pm 0.09$, while NuSeaC\cite{NuSeaC} finds at
$Q^2=7.4 $ GeV$^2$ that $\bar{U}-\bar{D}=-0.10\pm 0.024$.

Could this be due to the exclusion principle, as
illustrated in figure 5, which makes harder for $u$-type
antiquarks to appear in the sea? It could be, but it is
very difficult to come up with quantitative results.
A partial understanding is provided by the chiral quark model,
including pion exchange
(see figure 5), but this type of physics is still very poorly
understood.

\begin{figure}[h]
\epsfysize=2.5cm
\epsfbox{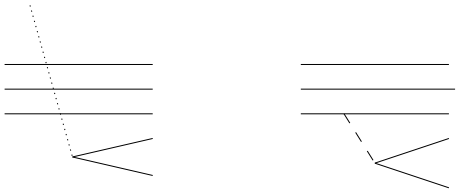}
\caption{Valence and sea quarks. Contribution from the chiral quark model.
\label{fig5}}
\end{figure}

\section{Scaling Violations}
It is plain clear from the data that there is
some $Q^2=-q^2$ dependence in the structure functions.
In other words, there are violations of Bjorken scaling
and actually $F_f=F_f(x,Q^2)$. The free parton model
is not completely correct (no big surprise, of course). Our job is to
try to understand these violations in the framework of
QCD.

Let us assume that we have isolated a parton with initial
momentum $p=xP$. The probability of finding such a parton
is given by $q(x)$. At the parton level the structure
function $F_2$ is just $\hat{F}_2= x \delta_f$ ($\delta_f$
is the appropriate charge. At the proton level, however,
this partonic cross-section has to be multiplied by
the probability of finding  the parton with momentum fraction $x$,
i.e. by $q(x)$. We write this in the form
\be
F_2(x)= x \delta_f \int_0^1 d\xi q(\xi)\delta(x-\xi)=
x\delta_f\int_0^1 \frac{d\xi}{\xi} q(\xi)\delta(1-\frac{x}{\xi}).
\label{parton1}
\ee

\begin{figure}[h]
\epsfysize=2.5cm
\epsfbox{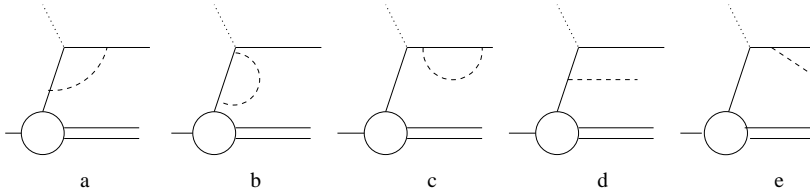}
\caption{Contribution at ${\cal O}(\alpha_s)$ to the relevant DIS
subprocess. \label{fig6}}
\end{figure}

At ${\cal O}(\alpha_s)$ many diagrams contribute. They are
given in figure 6. We are looking for scaling violations and,
therefore, we must look for logs. In other words we must investigate
ultraviolet, infrared and mass singularities of any kind. It turns out
that ultraviolet singularities are proportional to the
free result and simply renormalize $\delta_f$. Infrared
divergences cancel amongst all diagrams. Only the mass
singularity present in diagram (d) when the momentum of the gluon is
parallel to that of the gluon survives. Keeping only logarithmic terms,
the caculation at ${\cal O}(\alpha_s)$ amounts to the
replacement
\be
\delta(1-\frac{x}{\xi})\to \delta(1-\frac{x}{\xi})
+\frac{\alpha_s}{2\pi}P(\frac{x}{\xi})\log\frac{Q^2}{\lambda^2},\label{llog}
\ee
where $\lambda^2$ is an infrared regulator and
\be
P(z)=C_F[\frac{1+z^2}{(1-z)_+}+\frac{3}{2}\delta(1-z)].\label{splitt}
\ee
Only the logarithmic term is retained for this discussion.
Now we understand the reason for writing things in apparently
such a complicated way. First of all, scaling violations appear through
the contribution of real soft particles. $\xi$ is the original momentum
fraction of the nucleon carried by the parton,
which is reduced to $x\le \xi$ after the emission of the soft gluon. The
hard scattering takes place with the parton carrying fraction $x$.
All along the discussion, the transverse motion of the partons
inside the target is neglected as well as are all masses.

Then
\be
F_2(x)=
x\delta_f[q_0(x) + \int_x^1 \frac{d\xi}{\xi}
q_0(\xi)
\frac{\alpha_s}{2\pi}P(\frac{x}{\xi})\log\frac{Q^2}{\lambda^2}].\label{f2}
\ee
In addition we have replaced $q(x)$ by $q_0(x)$, the bare
PDF. If we now define the renormalized PDF by
\be
q(x,\mu^2)=q_0(x)+\frac{\alpha_s}{2\pi}\int_x^1 \frac{d\xi}{\xi}
q_0(x)P(\frac{x}{\xi})\log\frac{Q^2}{\lambda^2},\label{barepdf}
\ee
we can write
\be
F_2(x,Q^2)= x\delta_f[q(x,\mu^2)+\int_x^1 \frac{d\xi}{\xi}
q(x,\mu^2)\frac{\alpha_s}{2\pi}
P(\frac{x}{\xi})\log\frac{Q^2}{\mu^2}].\label{final}
\ee
No doubt the similarity with the usual renormalization
process did not go unnoticed. Now the infrared
regulator has been eliminated (hidden in the bare PDF)
at the expense of introducing a renormalization-scale dependence.
These are the sought after scaling violations.

\section{Altarelli-Parisi Equations and $\Lambda_{QCD}$}
At this point it is convenient to introduce the variable
$t=\frac{1}{2}\log \mu^2/\Lambda_{QCD}^2$. It then follows
from (\ref{barepdf}) that
\be
\frac{\partial}{\partial t} q(x,t) =
\frac{\alpha_s(t)}{\pi}\int_x^1 \frac{d\xi}{\xi} q(\xi,t)
P(\frac{x}{\xi}),\label{scaling}
\ee
which immediately translate into differential equations
for the structure functions themselves.
These are the Altarelli-Parisi equations\cite{alp}. They
summarize the rate of change of the parton distribution
functions with $t$.

We define the moments of the PDF's by
\be
q(n,t)=\int_0^1 dx x^{n-1} q(x,t). \label{mom1}
\ee
Introducing the anomalous dimension $\gamma_n$ as
\be
\gamma_n= \int_0^1 dx x^{n-1} P(x),\label{mom2}
\ee
the convolution over the fractional momentum $\xi$ transforms into
a product
\be
\frac{\partial}{\partial t} q(n,t) =
\frac{\alpha_s(t)}{\pi} \gamma_n q(n,t).\label{mom3}
\ee

This leads to following scaling behaviour for the moments of the
structure functions
\be
F_2(n,Q^2)=F_2(n,Q_0^2)
\left( {{\alpha_s(Q_0)}\over {\alpha_s(Q)}}
\right)^{{\gamma_{n}}\over {\beta_1}},\label{mom4}
\ee
which is our final expression. Experiments agree
on the whole very nicely with the scaling violations
predicted by QCD. Taking into account all the
subtle points of Quantum Field Theory that have gone
into the analysis, this provides a beautiful
check of the theoretical framework.

We have been considering $F_2$, but the same
procedure can be repeated for any structure function.
The expression (\ref{mom4}) amounts to resumming the leading
logs obtained by iteration of soft collinear
gluons. The diagram is the one shown in figure 7.

\begin{figure}[h]
\epsfysize=3cm
\epsfbox{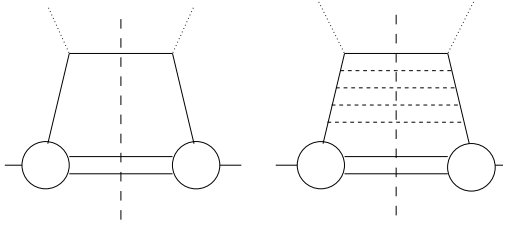}
\caption{Handbag and ladder diagrams.
\label{fig7}}
\end{figure}

As a simplifying hypothesis we have neglected mixing.
In fact, the evolution equation
is a $(2N_f+1)\times (2N_f+1)$ matrix, involving quarks
and gluons.
In the flavour singlet case life is
more complicated; there is mixing
with gluon operators and therefore one must also
consider gluon parton distribution functions as well
\be
{{\partial q(x,t)}\over {\partial t}}={{\alpha_s(t)}\over \pi}
\int_x^1 {{dy}\over y}[q(y,t)P_{qq}({x\over y})
+g(y,t) P_{gq }({x\over y})]\label{mixing1}
\ee
\be
{{\partial g(x,t)}\over {\partial t}}={{\alpha_s(t)}\over \pi}
\int_x^1 {{dy}\over y}[g(y,t)P_{gg}({x\over y})
+q(y,t) P_{qg }({x\over y})]\label{mixing2}
\ee
The detailed
form of the Altarelli-Parisi kernels at leading and NLO order can
be found in \cite{leading}.
No complete calculation exists yet
at the NNLO
to my knowledge, just some partial results.

It is important to realize that the Altarelli-Parisi equations
are not exact. They take into account the perturbative contribution
only (and this up to a given order in perturbation theory).
They also neglect transverse motion, which leads to corrections
of ${\cal O}(\Lambda_{QCD}/Q^2)$ to the leading results. These are
more easily dealt with in the perhaps more rigorous (but more cumbersome)
treatment based in the Operator Product Expansion\cite{ope}, which
will not be discussed here. Target mass corrections should be
equally taken into account. They are particularly important near
thresholds (such as the charm and bottom thresholds). For instance the
proper treatment of thresholds is highly relevant for HERA, since
a big chunk of the data comes from a region close to these thresholds.
And, of course, mass corrections are important for charm PDF from
the sea, which have actually been recently measured.

The analysis
of Deep Inelastic
Scattering
based on the Altarelli-Parisi
equations (or, alternatively, on the Operator Product Expansion)
has been one of the most clear tests of
perturbative QCD and traditionally the best way of determining $\alpha_s$,
which, as we have seen, enters in the scaling violations.
However, at present the value of $\alpha_s(M_Z)$
extracted from Z-physics is equally accurate if not more.

For some time a discrepancy was claimed between the value
of $\alpha_s$ obtained from  the analysis of scaling
violations and the Z-pole value. For instance
the value quoted by CCFR was
$\alpha_s(M_Z)=0.111\pm 0.004$, well below the world average\cite{pdg},
and quite away, for instance, from measurements based on event shapes
at the $Z$ peak ($\alpha_s(M_Z)=0.122\pm 0.007$; for measurements of
$\alpha_s$ at LEP, see e.g.\cite{lep}). The average $\alpha_s$ from DIS was
given\cite{average} just two years ago to be $\alpha_s(M_Z)=0.113\pm 0.005$.
Theoretical speculations
were fuelled and it was claimed that non-perturbative corrections
could be larger than originally thought.

Deep inelastic scattering data have been reanalyzed
recently and the quoted value for $\alpha_s(M_Z)$ from
DIS from a global fit\cite{global} to all data is $\alpha_s(M_Z)=0.118\pm
0.005$. The agreement with other determinations is now almost perfect.
The discrepancy was apparently due, it is claimed, to the energy calibration
of the detector (in the case of CCFR, at least; the new CCFR value is
$\alpha_s(M_Z)=0.119\pm 0.005$\cite{CCFR}),
a better understanding of higher twist corrections (not discussed here)
and from the treatment of the  contribution from heavy quarks from the sea,
charm in particular.
We will see later, when we discuss in somewhat more detail
the form of the PDF, that one must actually make
a number of hypothesis before being able to extract $\alpha_s(M_Z)$.
While do not regard the issue as totally settled yet, because
some of the older data sets are very poorly described by the
now preferred value of $\alpha_s(M_Z)$, the most
recent data coming from HERA\cite{hera} (H1 and ZEUS) give values
which agree nicely with each other and in
fact fall well in the high $\alpha_s$  range, almost on top of the new
average value. These ongoing experiments will become statistically more and
more significant in the near future in the determination
of $\alpha_s(M_Z)$ as more and  more data points pile up.

\section{Parton Distribution Functions}
We do not know
in general how to compute the
parton distribution functions, even for $-q^2\to\infty$.
Only their evolution can be reliably computed either
through the Operator Product Expansion of the use
of the Altarelli-Parisi equations and this for
large enough values of $-q^2$.
The scaling behaviour is governed by the
anomalous dimensions. At leading order they are
\be
\gamma_{qq}(j)= C_F[-\frac{1}{2}+\frac{1}{j(j+1)}
-2\sum_{k=2}^j\frac{1}{k}],\label{ano1}
\ee
\be
\gamma_{qg}(j)=T_R[\frac{2+j+j^2}{j(j+1)(j+2)}],\label{ano2}
\ee
\be
\gamma_{gq}(j)=C_F[\frac{2+j+j^2}{j(j^2-1)}],\label{ano3}
\ee
\be
\gamma_{gg}(j)=2C_A[-\frac{1}{12}+\frac{1}{j-1}
+\frac{1}{(j+1)(j+2)}-\sum_{k=2}^j\frac{1}{k}]-\frac{2N_f}{3}T_R,\label{ano4}
\ee
where $C_F,T_R$ and $C_A$ are group-theoretical factors.

An interesting issue is the behaviour
of the parton distribution functions at the
endpoints $x=0$ and $x=1$. The large
$n$ behaviour of the moments probes
the $x\to 1$ region.
Since it is natural to expect that
at the kinematical boundaries the
parton distribution functions vanish,
one can make the following ansatz
for $x\to 1$
\be
q(x,Q^2)\sim A(Q^2) (1-x)^{\nu(\alpha_s(Q^2))-1}.\label{ansatz1}
\ee
Demanding that eq. (\ref{ansatz1}) fulfills the $q^2$ evolution
equation  leads to
\be
A(Q^2)=A_0 {{[\alpha_s(Q^2)]^{-d_0}}\over
{\Gamma(1+\nu(\alpha_s(Q^2)))}}\qquad
\nu(\alpha_s)=\nu_0 -{{16}\over {33-2N_f}}\log\alpha_s(Q^2),
\label{ansatz2}
\ee
\be
d_0={{16}\over {33-2N_f}}({3\over 4}-\gamma_E).\label{ansatz3}
\ee
Likewise, for the gluons we have
\be
g(x,Q^2)\sim A_0^\prime {{[\alpha_s(Q^2)]^{-d_0}}\over
{\Gamma(2+\nu(\alpha_s(Q^2)))}}
{{(1-x)^{\nu(\alpha_s(Q^2))}}\over {\log(1-x)}}.\label{ansatz4}
\ee
The constants $A_0, A^\prime_0$ and $\nu_0$ are not
calculable on perturbative QCD and depend on the
specific operator. $d_0$ is universal.

When $x\to 1$ the gluon distribution functions
approach zero more rapidly than the quark ones. For
large values of $x$ the quark contents of nucleons
is the relevant one. Second order corrections to
this asymptotic behaviour can be derived in a similar way and are known.
It turns out that the correction is arbitraryly large if one
gets sufficiently close to $x=1$. This is because the collinear
gluon is, in addition, soft in that exceptional configuration thus giving
rise to a $\log(1-x)$ singularity. Multiple emission
is then kinematically favoured, since the log overcomes
the $\alpha_s$ suppression.

For small values of $x$ the
opposite behaviour takes place, the gluon distribution
function eventually becomes dominant. At
LHC the cross-section
will be greatly dominated by low-$x$ physics and the
important process there will be gluon-gluon scattering. At the current
Tevatron run the quark contents of protons and antiprotons is still
dominant. Let us see why gluons dominate completely
at low $x$.

The key point is the appeareance of the singularity for $j=1$.
Indeed, as $j\to 1$
\be
\gamma_{gg}(j)\sim \frac{2N}{j-1}.
\ee
Then
\be
g(j,t)=g(j,t_0)\exp[\frac{N}{\pi\beta_1(j-1)}\log\frac{t}{t_0}].
\ee
Then
we
 proceed to evaluate $g(x,t)$ by performing an inverse Mellin transform
\be
xg(x,t)=\frac{1}{2\pi i}\int_C dj x^{1-j} g(j,t).
\ee
The integration circuit is a line in the direction of the
imaginary axis in the complex $j$ plane, to the right of the $j=1$
singularity.
The saddle point method can now be used provided
that $\log(1/x)$ is large and
this is the reason why this procedure gives only the
small $x$ behaviour. Working things out we see that the
gluon parton distribution function for low
$x$ behaves as
\be
g(x)\sim {1\over x}\exp\sqrt{C(Q^2)\log{1\over x}},\label{gluonx}
\ee
where $C(Q^2)$ is calculable. Unfortunately, this answer is
not totally satisfactory because something must stop the
growth in $g(x)$ for low $x$, or else one runs into unitarity problems
sooner or later, and thus eq. (\ref{gluonx}) it is not credible all the way
to $x=0$. Technically speaking, there must be corrections that destabilize
the saddle point solution. Physically, the uncontrolled growth of the
gluon distribution is an infrared unstability. The density of soft gluons
is too large. Shadowing and non-linear evolution equations are
the buzzwords here\cite{shadowing}.

The double scaling limit (high $Q^2$, low $x$) is well
supported by the data. See for instance\cite{doublescal} for a
recent analysis coming from measurements of $F_2$ at H1.

Except in these two limiting cases, PDF's have to be parametrized.
The way one proceeds is by proposing a given parametrization
at some reference value $Q_0^2$, then evolve to all desired values
of $Q^2$ using the Altarelli-Parisi equations, then perform a global fit
of the parameters describing the PDF and, at the same time, determine
$\alpha_s$.

\section{Confinement}
$\Lambda_{QCD}$ sets a natural scale in
the theory. Well above $\Lambda_{QCD}$ perturbation
theory makes sense.  Of course perturbative QCD at
large enough energies describes a world
of quasi-free quarks, interacting with
Coulomb-like forces. We know very well
that hadronic physics is a very different
world where quarks are confined into
colorless hadrons.
As soon as
$q^2\sim\Lambda_{QCD}^2$ perturbation theory is
unreliable. It simply cannot explain confinement.

What confinement means is that there is a force
between quarks that does not decrease with distance.
There is indeed phenomenological evidence
(which is supported by lattice analysis) that the
interquark potential at large distances in QCD is of the
form
\be
V(r)\sim a\Lambda^2_{QCD}r -{b\over r} +\dots
\label{conf}
\ee
The first term is a confining quark potential. The
constant $a$ has to be  $\sim 1$ because $\Lambda^2_{QCD}$
 is the only dimensional quantity at our disposal.
The Coulombic part is called the L\"uscher term and plays
a crucial role in heavy quark spectroscopy\cite{hquark}.

At short distances the behaviour of the interquark potential
is totally different. It is Coulomb-like
\be
V(r)\sim -\frac{\alpha_s}{r}.\label{coul}
\ee

Different quarks probe the different regimes.
Indeed, because  $\alpha_s(m_t)$ is
so small (say $\sim 0.1$), the corresponding Bohr radius
is $r_0\sim 10^{-2}$ fm, much
smaller than $\Lambda_{QCD}^{-1}$. The
coulombic part of the interquark potential
largely dominates. (At such short
distances the linearly rising potential
is not at work, the leading confinement
effects are $\sim r^3$, as discussed
by Leutwyler some time ago\cite{leut}, but they can
be safely neglected at first approximation.)

Bottom and charm are in a somewhat intermediate
position. $\alpha_s(m_b)$ is still
relatively small. The Bohr radius
is $10^{-1}$ fm, smaller but comparable
to $\Lambda_{QCD}^{-1}$. Spectroscopy
is basically perturbative, at least for the lowest levels, but some
non-perturbative effects are visible. Charm is really no-man's land. Both
perturbative and non-perturbative effects compete even for the
ground state $n=1$.
For light quarks the Bohr radius is several fm and the confining
potential is fully at work.

The existence of a confining potential leads
to very large
multiplicities and jets.
One can imagine
a quark-antiquark being formed at the
primary vertex then moving apart. Part of
their kinetic energy is deposited in the interquark
potential as they move away. Very quickly
a separation $r_m$ is reached where the energy deposited
is enough to form a new quark-antiquark pair,
\be
\Lambda^2_{QCD} r_m\simeq 2 m_q,\label{break}
\ee
at that moment the quark-antiquark `string'
breaks and the process is repeated until
the average relative momentum is small
enough and hadronization takes
place.

There is a lot of physics in the string picture. We can think
of color forces being confined in some sort of tube or string joining
the two moving quarks. The chromodynamic energy is thus stored in
a relatively small region of space-time. If this picture is correct
we should expect hadronization to take place in this region in
preference to any other. This is indeed the case; in three jet events
(which originate from $\bar{q} q g$, with a hard gluon) there is a clear
enhancement
of soft gluon and hadron production in the regions between color
lines (representing the gluon by a double color line, or $\bar{q} q$
state), and a relative depletion in other regions.
This phenomenon is called color coherence\cite{cohere}.

\section{Dual Models}
We now backtrack in history, to the pre-QCD days, and recall
that in the 60's the duality hypothesis was much in fashion.
The hypothesis stated that in strong interactions
a sum over intermediate states in the $s$-channel should
reproduce the sum over resonances in the $t$-channel.
Mathematically,
\be
A(s,t)=\sum_J \frac{g_J^2 s^J}{t-M_J^2}=
\sum_J\frac{g_J^2t^J}{s-M_J^2}.\label{duality}
\ee
Of course for this to have even a chance of being true an infinite
number of intermediate states is required. It should be
stated right away that the evidence for this peculiar
property was (and still is) rather weak.

However, in 1968, Veneziano\cite{venezia} took the
idea seriously and proposed the following amplitude
\be
A(s,t)=\frac{\Gamma(-\alpha(s))\Gamma(-\alpha(t))}{
\Gamma(-\alpha(s)-\alpha(t))}, \qquad \alpha(s)=\alpha(0)+
\alpha^\prime s.\label{venamp}
\ee
This amplitude is manifestly dual. In 1969 and 1970 Y.Nambu and
others unveiled the relation between the Veneziano amplitude and
open string theory. It was later generalized to closed strings
by Koba and Nielsen.

This of course is the way to make contact
with the long-distance properties of QCD that we have discussed
in the previous section. If in some kinematical regime
QCD can be described by some type of string theory, an amplitude
of the Veneziano type should describe strong interactions
in a regime where perturbation theory is not valid.

Unfortunately life is not so easy. First of all, consistency
of string theory requires $\alpha(0)=1$ and then the Veneziano
amplitude exhibits poles in the $s$-channel
whenever $s=(n-1)/\alpha^\prime$, $n=0,1,...$. There is
a tachyonic scalar/pseudoscalar particle and a massless vector particle.
In addition, the amplitude does not exhibit the
proper chiral behaviour (Adler zero), i.e. $A(0,0)=0$, if
we are to interpret the pseudoscalars as pions. It is
a complete phenomenological failure.

Lovelace and Shapiro\cite{lovelace} finally proposed an amplitude
with the correct behaviour or, at least, not manifestly
incorrect. It is inspired from the theory
of supersymmetric strings  and the equivalent of the
Veneziano amplitude now reads
\be
A(s,t)=\frac{\Gamma(1-\alpha(s))\Gamma(1-\alpha(t))}
{\Gamma(1-\alpha(s)-\alpha(t))},\qquad \alpha(s)=\alpha(0)+\alpha^\prime s,
\label{shapiro}
\ee
where
$\alpha(0)$ in principle also equals 1 here. Again this amplitude (which
is tachyon-free --- this is the main virtue of the supersymmetric string) is
not physically acceptable because the amplitude does not have the correct
Adler zero. However, it does reproduce the right chiral behaviour
if one replaces by hand this value by 1/2 and is then appropriate to
describe pion scattering. Unfortunately this replacement cannot
be derived at present from any known string theory. The corresponding
trajectory is called the Regge trajectory and corresponds to the
exchange of open strings and, as discussed, has an intercept $\alpha_R=1/2$.
Physically this is interpreted as the exchange of quark-antiquark pairs
In addition there is a Pomeron trajectory
which has an intercept $\alpha_P=\alpha_R+1/2$, and is due to the
exchange of closed strings
(interpreted as glueballs in QCD).

\begin{figure}[h]
\epsfysize=2cm
\epsfbox{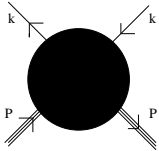}
\caption{Nucleon-parton scattering process.
\label{fig8}}
\end{figure}

Let us now return to DIS and let us see what all this has to do
with it. Let us consider the behaviour of the
amplitude $A(s,t)$ for large $s$ and fixed $t$
\be
A(s,t)\sim s^{\alpha^\prime t + \alpha_0}. \label{larges}
\ee
Let us now assume that the elastic nucleon-parton amplitude (figure 9)
is described by such an amplitude. Kinematically, $s=(P-k)^2$. If we
decompose
\be
k=xP-\frac{k_T^2}{2x}n + k_T,\label{decomp}
\ee
where $n$ is a vector such that
$n^2=0$, $n\cdot P=1$ and $k_T\cdot n= k_T\cdot P = 0$, then, with
the usual approximations,
\be
s=-2k\cdot P = -\frac{k_T^2}{x}\label{relat}
\ee
We see then that, at a finite value of $k_T$, low $x$ corresponds to the
large $s$ behaviour. On the other hand, for this subprocess $t=0$, and
we of course realize that the amplitude is directly related to
the cross-section for parton + proton $\to $ anything.
If the parton is a quark
only the Reggeon will contribute. If there is mixing with partonic gluons
we will have a contribution from both the Regge and Pomeron trajectories.
After a short calculation we shall conclude that
\be
 F_2(x)\sim A_P x^0 + A_R x^{\frac{1}{2}}.\label{predict}
\ee
And, therefore that, at low values of $x$,
\be
 g(x)\sim x^{-1},\qquad q(x)\sim x^{-\frac{1}{2}}.
\ee
These are the implications of Regge theory for the parton
distribution functions. Notice that there is no $Q^2$
dependence anywhere, so the question poses itself
as to which is the appropriate value of $Q^2$ to compare with.
The answer is not very well defined, but it should
correspond to the typical range of energies where Regge
phenomenology is known to be valid in other contexts, i.e.
a few GeV. In fact, a fit to the gluon PDF
shows that at $Q^2 =4$ GeV$^2$, $g(x)\sim x^{-1.17}$.
Not bad.

\section{Low $x$ region}
The region
below $10^{-2}$
had not been explored experimentally until very recently;
a first look at these low-$x$
values has been provided by the commissioning of HERA.
HERA is a machine ideally suited for an in-depth analysis
of structure functions.
It should be possible to arrive at very low values
of $x$ (down to $x\sim 10^{-5}$).

Most parametrizations have traditionally performed very poorly
when extrapolated to the low $x$ region.
Typically they predict an increase as
$x\to 0$ which is lower than what is actually seen.
The behaviour $F_2(x)\sim x^{-\lambda}$, with
$\lambda\sim 1/2$ as $x\to 0$, which is predicted from the
BFKL evolution equation\cite{bfkl} seemed at some point (see
e.g.\cite{martin})
to stand the comparison with HERA results best. However, this behaviour is
still incompatible with unitarity and cannot hold all the way
to $x=0$ either. If fact we know now that the predictions from
BFKL cannot be trusted. This has prompted a renewed interest
in trying to extract the behaviour at low $x$ from
conventional Altarelli-Parisi evolution. The consensus
now seems to be that even for the
low values of $x$ analyzed at HERA there is no real evidence
of any results beyond ordinary perturbative QCD.

It is easy to understand why perturbative QCD must fail at some
point. The expansion of the splitting function
$P(z)$ in powers of $\alpha_s$ at the NLO actually resums
all terms of the form  $(\alpha_s \log Q^2)^n$ and
$\alpha_s^n\log^{n-1} Q^2$. Looking at (\ref{propa})
we see that the propagator causing the mass singularity is
($p=\xi P$)
\be
\frac{1}{2pk}=-\frac{2x}{\xi k_T^2}.\label{kate}
\ee
Apart from the parametric integrals, we have
\be
\int \frac{d^2 k_T}{k_T^2}.\label{corner}
\ee
This is the origin of the $\log\lambda^2$ and, eventually,
of the $\log Q^2$.

The leading $\log^n Q^2$ will thus be produced
by one single region in integration
\be
\int \frac{d^2 k_T^n}{(k_T^n)^2}
  \int \frac{d^2 k_T^{n-1}}{(k_T^{n-1})^2} ... \int \frac{d^2
k_T^1}{(k_T^1)^2},\label{corner1}
\ee
with $\vert Q\vert \gg \vert k_T^n\vert\gg\vert k_T^{n-1}\vert\gg
... \vert k_T^1\vert $.

At sufficiently large $x$ logarithms of $1/x$ necessarily appear.
We have actually seen them in the double scaling limit. They must,
at some point, spoil the predictivity of the
perturbative expansion. One must then identify the regions
in integration capable of giving rise to terms
of the form
$(\alpha_s \log\frac{1}{x})^n$,
and eventually to $\alpha_s^n \log^{n-1}\frac{1}{x}$ and so on.

\begin{figure}[h]
\epsfysize=4cm
\epsfbox{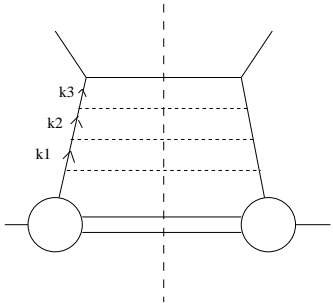}
\caption{Ordering leading to the most singular $\log{1/x}$ contribution.
\label{fig9}}
\end{figure}

Lipatov and coworkers\cite{bfkl} (see also\cite{stefano} for alternative
derivations) have identified such a contribution.
It corresponds to the diagram depicted in figure 10, more specifically to the
region
\be
k_i=\alpha_iP + \beta_i n + K_{i T},\label{bfkl1}
\ee
\be
\alpha_1\gg \alpha_2\gg ...\gg \alpha_{n-1},\qquad
k_{i T}\sim k_{j T},\qquad
\beta_1\ll \beta_2\ll ... \ll \beta_{n-1}.\label{bfkl3}
\ee
This leads to splitting kernels similar to those
of the Altarelli-Parisi equations
\be
F_2(x,Q^2)=\int d^2 k_T\int_x^1 \frac{d\xi}{\xi}
C(\frac{x}{\xi},Q^2,k_T) F_2(\xi,k_T),\label{bfkl4}
\ee
where $ F_2(\xi,k_T)$ obeys the differential equation
\be
\xi\frac{\partial}{\partial\xi} F_2(\xi,k_T^\prime)=
\int d^2k_T K(k_T^\prime,k_T) \frac{{k^\prime_T}^2}{k_T^2}
F_2(\xi, k_T).\label{bfkl5}
\ee
The BFKL kernel is now known to leading and subleading
order.\cite{bfkl,bfkl2} The leading asymptotic solution is
\be
F(x,k_T)\sim x^{-4N \log\frac{2\alpha_s}{\pi}}.\label{bfkl6}
\ee
Unfortunately the corrections implied by the next-to-leading
calculations are gigantic\cite{bfkl2}. There is no way of doing
anything useful with BFKL scaling at present.

As previously discussed this does not seem to be a problem for
HERA data since a careful analysis shows that ---perhaps surprisingly---
the data is well accounted for by ordinary perturbative QCD (the matter
is however somewhat controversial to this date), but it will
come the day where $\log 1/x$ corrections will be essential. The subject is
thus still open.

\section*{Acknowledgements}
It is a pleasure to thank S.Forte for discussions on some of
the topics presented in this lecture. This work has been
supported in part by CICYT grant AEN98-0431 and
CIRIT project 1998SGR 00026. Being at the meeting has
been, as always, a great pleasure. Thanks are due to B.Adeva
and the rest of the organizing committee for the invitation
extended to me. Thanks are also due to Mercedes Fatas, the
efficient secretary of the XXVI International Meeting on Fundamental Physics
for her patience in dealing with this
long overdue contributor to the proceedings.

\end{document}